\documentclass{moriond}

\usepackage{amssymb,amsmath,amsfonts,graphicx,mathrsfs,color,bm}
\usepackage{enumerate}
\usepackage{empheq}
\usepackage{tikz}
\usepackage{tikz-cd}
\usepackage{physics}

\def\bea{\begin{eqnarray}}
\def\eea{\end{eqnarray}}

\newcommand{\be}{\begin{equation}}
\newcommand{\ee}{\end{equation}}
\newcommand{\sbe}{\begin{subequations}}
\newcommand{\see}{\end{subequations}}

\newcommand{\ba}{\begin{eqnarray}}
\newcommand{\ea}{\end{eqnarray}}
\newcommand{\p}{\partial}
\newcommand{\nn}{\nonumber}

\newcommand{\di}{\mathrm{i}} 
\newcommand{\de}{\mathrm{e}}
\newcommand\calO{{\mathcal{O}}}

\allowdisplaybreaks

\newcommand{\bigO}[1]{\mathcal{O}\left(\frac{1}{c^{#1}}\right)}

\newcommand{\Photo}{}

\begin{document}
\vspace*{4cm}
\title{GRAVITATIONAL WAVES IN SCALAR-TENSOR THEORY TO ONE-AND-A-HALF POST-NEWTONIAN ORDER}

\author{David TRESTINI}
\address{GReCO, Institut d’Astrophysique de Paris, UMR 7095, CNRS \& Sorbonne Université, 98bis boulevard Arago, 75014 Paris, France\\
Laboratoire Univers et Théories, Observatoire de Paris, Université PSL, Université de Paris, CNRS, F-92190 Meudon, France}


\maketitle\abstracts{
We compute the gravitational waves generated by compact binary systems in a class of massless scalar-tensor (ST) theories to the 1.5 post-Newtonian (1.5PN) order beyond the standard quadrupole radiation in general relativity (GR). Using and adapting to ST theories the multipolar-post-Minkowskian and post-Newtonian formalisms originally defined in GR, we obtain the tail and non-linear memory terms associated with the dipole radiation in ST theory. The multipole moments and GW flux of compact binaries are derived for general orbits including the new 1.5PN contribution, and comparison is made with previous results in the literature. In the case of quasi-circular orbits, we present ready-to-use templates for the data analysis of detectors, and for the first time the scalar GW modes for comparisons with numerical relativity results.}

\section{Massless scalar-tensor theories}\label{sec:STtheory}


We consider a generic class of scalar-tensor theories in which a single massless scalar field $\phi$ minimally couples to the metric $g_{\mu\nu}$. It is described by the skeletonized action for $N$ spinless point-particles \cite{Mirshekari:2013vb}
\be\label{STactionJF}
S = \frac{c^{3}}{16\pi G} \int\dd^{4}x\,\sqrt{-g}\left[\phi R - \frac{\omega(\phi)}{\phi}g^{\alpha\beta}\p_{\alpha}\phi\p_{\beta}\phi\right]  - c \sum_{A} \int\,m_{A}(\phi) \sqrt{-\left(g_{\alpha\beta}\right)_{A}\dd y_{A}^{\alpha}\,\dd y_{A}^{\beta}}\,,
\ee
where $R$ and $g$ are respectively the Ricci scalar and the determinant of the metric, $\omega$ is a function of the scalar field, $y_A^\alpha$ denote the space-time positions of the particles, and $\left(g_{\alpha\beta}\right)_{A}$ is the metric evaluated at the position of particle~$A$.  A major difference in ST theories compared to GR is that, as a consequence of the breaking of the strong equivalence principle, we have to take into account the internal gravity of each body by introducing a scalar-field-dependant mass $m_A(\phi)$~\cite{Eardley1975}. 

To go from the ``Jordan frame'' to ``Einstein frame'', we define a rescaled scalar field and the conformally related metric as
\be
\varphi\equiv \frac{\phi}{\phi_{0}}\,,\qquad\qquad\tilde{g}_{\alpha\beta}\equiv \varphi\,g_{\alpha\beta}\,,
\ee
so that the physical and conformal metrics have the same asymptotic behaviour at spatial infinity. The scalar field is then minimally coupled to the conformal metric, so we will perform most of our computations using this conformal metric and go back to the physical metric only in the end when computing observable quantities. 

Next, we define the scalar and metric perturbation variables  $\psi\equiv\varphi-1$ and $h^{\mu\nu}~\equiv~ \sqrt{-\tilde{g}}\tilde{g}^{\mu\nu}~-~\eta^{\mu\nu}$ where $\eta^{\mu\nu}~=~\text{diag}(-1,1,1,1)$ is the Minkowski metric. We impose the harmonic gauge condition on the conformal metric : $\partial_\mu h^{\mu\nu} = 0$. The field equations derived from the action~\eqref{STactionJF} then read
\begin{subequations}\label{rEFE}
\begin{align}
& \Box_{\eta}\,h^{\mu\nu} = \frac{16\pi G}{c^{4}}\left( \frac{\varphi}{\phi_{0}} \vert g\vert T^{\mu\nu} +\frac{c^{4}}{16\pi G}\Lambda^{\mu\nu} \right)\,,\\
& \Box_{\eta}\,\psi = \frac{8\pi G}{c^{4}}\left(\frac{\varphi}{\phi_{0}(3+2\omega)}\sqrt{-g}\left(T-2\varphi\frac{\p T}{\p \varphi}\right) -\frac{c^{4}}{8\pi G}\Lambda_s\right)\,,
\end{align}
\end{subequations}
where $\Box_{\eta}$ denotes the ordinary flat space-time d'Alembertian operator. Here $T^{\mu\nu}= 2 (-g)^{-1/2}\delta S_{\mathrm{m}}/\delta g_{\mu\nu}$ is the matter stress-energy tensor, $T\equiv g_{\mu\nu}T^{\mu\nu}$ and $\p T/\p \varphi$ is defined as the partial derivative of $T(g_{\mu\nu}, \varphi)$ holding $g_{\mu\nu}$ constant. The non-linearities in the scalar source read
\be\label{lambdas} \Lambda_s = -h^{\alpha\beta}\p_{\alpha}\p_{\beta}\psi-\p_{\alpha}\psi\p_{\beta}h^{\alpha\beta} +\left(\frac{1}{\varphi}-\frac{\phi_{0}\omega'(\phi)}{3+2\omega(\phi)}\right)\tilde{\mathfrak{g}}^{\alpha\beta}\p_{\alpha}\psi\p_{\beta}\psi\,. 
\ee 
We write the tensor source as $\Lambda^{\mu\nu} =\Lambda_{\mathrm{GR}}^{\mu\nu}+\Lambda_{\phi}^{\mu\nu}$, where  $\Lambda_{\mathrm{GR}}^{\mu\nu}$ is given by Blanchet \cite{Blanchet:2013haa}, where $h$ is now defined as in Eq. \ref{rEFE} and $\Lambda_{\phi}^{\mu\nu}$ is given by
\label{exprLambda}
\begin{align}\Lambda_{\phi}^{\mu\nu} = \frac{3+2\omega(\phi)}{\varphi^2}\left(\tilde{\mathfrak{g}}^{\mu\alpha}\tilde{\mathfrak{g}}^{\nu\beta} -\frac{1}{2}\tilde{\mathfrak{g}}^{\mu\nu}\tilde{\mathfrak{g}}^{\alpha\beta}\right)\p_{\alpha}\psi\p_{\beta}\psi\,.
\end{align}

\section{The Multipolar post-Minkowskian formalism in scalar-tensor theories}\label{sec:MPNinST}

\subsection{The scalar-tensor multipole moments}
In order to solve the vacuum field equations $\Box h^{\mu\nu} = \Lambda^{\mu\nu}$ and $\Box \psi = \Lambda_s$ in the exterior region of the isolated matter system, we perform a multipolar decomposition (indicated by $\mathcal{M}$) conjointly with a non-linear post-Minkowskian expansion~\cite{Blanchet:1985sp}. In the exterior region we can formally write the multipolar-post-Minkowskian expansion of the tensor and scalar field as
\begin{subequations}
\label{PMexpansion} 
\begin{align}
 		h^{\mu\nu}_{\mathrm{ext}} &\equiv \mathcal{M}(h^{\mu\nu}) = G h_1^{\mu\nu} + G^2 h_2^{\mu\nu} + \calO\left( G^3\right)\,,\label{h1h2}\\
		\psi_{\mathrm{ext}} &\equiv \mathcal{M}(\psi) = G \psi_1 + G^2 \psi_2 + \calO\left(G^3\right)\,.
\end{align}
\end{subequations}
To obtain the expressions of the source and current moments $I_L$ and $J_L$ (relative to the tensor sector) as functions of the source, the procedure is essentially the same as in GR, with the caveat that the moments are defined with respect to the conformal metric. The multipole expansion of $h^{\mu\nu}$ is obtained using a matching to the PN expansion in the near zone of the source~\cite{Blanchet:1998in}. In addition to $I_L$ and $J_L$, we introduce a new set of scalar multipole moments which we call $I_L^s$, also chosen to be symmetric-trace-free (STF). We find
\begin{subequations}
\begin{align} 
\mathcal{M}(\psi) &= \mathop{\mathrm{FP}}_{B=0}\,
\Box^{-1}_\mathrm{ret} \Bigl[ \left(\frac{r}{r_0}\right)^B \mathcal{M}(\Lambda^s)\Bigr] + G \psi_1 \,,\label{Mpsi}\\
\text{with}\quad\psi_1 &= - \frac{2}{c^2}
\sum_{\ell=0}^{+\infty}
\frac{(-)^\ell}{\ell!}\,\partial_L\!\left[\frac{I_L^s(u)}{r}\right]\,.\label{psi1}
\end{align}
\end{subequations}
Similarly to the GR case, the scalar moments are obtained in closed form and we find that
\be\label{psi1mult} 
I^s_L(u) = \mathop{\mathrm{FP}}_{B=0} \int \dd^3 \mathbf{x}\,\tilde{r}^B  \int_{-1}^1 \dd z \,\delta_\ell(z) \,\hat{x}_L \Sigma^s ( \mathbf{x}, u+z r/c)\,.
\ee
where $$\Sigma_s \equiv  \frac{\varphi}{\phi_{0}(3+2\omega)}\sqrt{-g}\left(T-2\varphi\frac{\partial T}{\partial \varphi}\right)  -\frac{c^4}{8\pi G}\Lambda_s $$
Note that in contrast to the tensor monopole $I$, the scalar monopole $I^s$ is not constant but its time-variation will be a post-Newtonian effect, \textit{i.e.} $\dd I^s/\dd t = \mathcal{O}(c^{-2})$. 

\subsection{Non-linear effects in ST theory}

Once the vacuum linearized solutions for scalar \eqref{psi1} and tensor fields are obtained, with explicit expressions for the multipole moments as integrals over the PN expansions of $\tau^{\mu\nu}$ and $\tau_s$, the non-linear contributions can be computed by adapting to ST theories the Multipolar-post-Minkowskian (MPM) algorithm~\cite{Blanchet:1985sp}.

The equations to be solved for the quadratic metric are
\begin{subequations}\label{eqordre2}
\begin{align}
 \Box h^{\mu\nu}_2&= \Lambda^{\mu\nu}_{\text{GR},2}(h_1,h_1) + \Lambda^{\mu\nu}_{\phi,2}(\psi_1,\psi_1)\,,\\
  \Box \psi_2&= \Lambda^{(h\times\psi)}_{\text{S},2}(h_1,\psi_1) + \Lambda^{(\psi\times\psi)}_{\text{S},2}(\psi_1,\psi_1)  \,,
\end{align}
\end{subequations}
where the source terms are reduced to their quadratic contributions with respect to $\psi_1$ and~$h_1$. 

Following the MPM algorithm, we can solve these equations and find a solution valid in the exterior zone and in harmonic gauge. However, its expression contains logarithms at spatial infinity. Of course, this is only a coordinate effect, which can be cured by introducing a radiative type coordinate system $(T, R)$, with $U \equiv T - R/c$ being an asymptotically null coordinate such that $U = u - \frac{2 G I}{c^3}\ln\left(\frac{r}{c b}\right) + \mathcal{O}\left(\frac{1}{r}\right)$
where $I$ is the mass monopole moment and $b$ is an arbitrary constant time-scale.

\subsection{Some non-local effects: tails and memory}

At 1.5PN order, the non-linear metric includes non-local tail and memory terms. Tails can be interpreted as linear waves reflecting against the curvature of space-time: the lowest order tail effect is a quadrupole-mass interaction for gravitational waves and a dipole-mass interaction for a scalar wave. Conversely, memory effects can be seen as linear waves generated by linear waves. In our case, it is a scalar dipole-dipole interaction that enters the gravitational waveform. The non-local expressions for the tail and memory terms that enter the quadratic waveform are:
\begin{subequations}\label{tailsmemory}
\begin{align}
h^{ij}_{2,\text{TT}} &=  - \frac{2G^2}{c^7 R}\perp^\text{TT}_{ijab}\Bigg\{\frac{2 M}{\phi_0} \int_{-\infty}^U \!\dd V \overset{(4)}{I_{ij}}(V) \left[ \ln  {\left(\frac{U-V}{2b} \right)} + \frac{11}{12} \right]+ \frac{3+2\omega_0}{3}\int_{-\infty}^U \!\dd V \overset{(2)}{I^s_{\langle i}}(V)\overset{(2)}{I^s_{j\rangle}}(V)\Bigg\}\nn\\
&\qquad+ \text{(instantaneous terms)} + \bigO8 \,, \\
\psi_ 2 &= -\frac{4 G^2 M}{c^6 R \phi_0}\Bigg\{N_i  \int_{-\infty}^U \dd V \overset{(3)}{I_i^s}(V) \left[ \ln{\left(\frac{U-V}{2b} \right)} + 1 \right] + \frac{1}{c \phi_0 } \int_{-\infty}^U \dd V \overset{(2)}{E^s}(V)\ln\left(\frac{U-V}{2b}\right) \nn\\
& \qquad\qquad\qquad+\frac{N_{ij}  }{2c} \int_{-\infty}^U \dd V \overset{(4)}{I^s_{ij}}(V)\left[ \ln{\left(\frac{U-V}{2b} \right)}\right]  \Bigg\} + \text{(instantaneous terms)} + \bigO8 \,,
\end{align}
\end{subequations}

where $F^{(n)}(t) \equiv \dd^n F / \dd t^n$,  $\perp^\text{TT}_{ijab}$ is the transverse-traceless (TT) projector, $M\equiv I/\phi_0$ and $E_s$ is defined such that $\dd E_s / \dd t \equiv c^2\phi_0\, \dd I_s / \dd t= \mathcal{O}(1)$, since\,  $\dd I^s/\dd t = \mathcal{O}(c^{-2})$.

\section{Results}

\subsection{Spherical harmonic decomposition of the fields}

For quasi-circular orbits, we introduce the dimensionless PN variable $x \equiv (\tilde{G} \alpha m \omega/c^3)^{2/3}$, thanks to which the tail and memory terms can be integrated explicitly. The expressions for $h^{\mu\nu}$ and $\psi$ simplify considerably, and we can easily decompose the tensor field on its ``plus'' and ``cross'' modes, and perform the following spherical harmonic expansion for the tensor and scalar modes

\be h_+ - \di h_- = \sum_{\ell=2}^{+\infty} \sum_{m=-\ell}^\ell h^{\ell m} \,_ {-2}Y^{\ell m}\qquad\text{and}\qquad\quad  \psi =\sum_{\ell=0}^{+\infty} \sum_{m=-\ell}^\ell \psi^{\ell m}\  Y^{\ell m}\ee
where we have used spin-weighted spherical harmonics of weight -2 and 0. We then define the phase variable $\psi$ so as to remove most of the logarithms from the expressions of $h^{\ell m}$ and $\psi^{\ell m}$, which can always be written as
\begin{subequations}
\be
h^{\ell m} = \frac{2 \tilde{G}(1-\zeta) m \nu x}{R c^2}  \sqrt{\frac{16\pi}{5}} \,\hat{H}^{\ell m} \de^{-\di m \psi } \qquad\text{and}\qquad
\psi^{\ell m} = \frac{2 i \tilde{G} \zeta \sqrt{\alpha}\mathcal{S}_{-} m \nu \sqrt{x}}{R c^2} \sqrt{\frac{8\pi}{3}} \,\hat{\Psi}^{\ell m} \de^{-\di m \psi}\,,
\ee
\end{subequations}
with normalized modes $\hat{H}^{\ell m}$ and $\hat{\Psi}^{\ell m}$ defined such that $\hat{H}^{22} = 1 + \mathcal{O}(x)$ and $\hat{\Psi}^{11} = 1 + \mathcal{O}({x})$.
 
In this work, we computed all relevant tensor and scalar modes to 1.5PN order. The scalar modes had never been computed, and the dominant $\psi^{11}$ mode is given by :
 
\begin{align}
\hat{\Psi}^{11}&= 
1 + x \Bigg\{- \frac{9}{5} -  \frac{2}{3} \bar{\beta}_{+} -  \frac{1}{3} \bar{\gamma} + 2 \bar{\beta}_{+} \bar{\gamma}^{-1} + \frac{2 \bar{\beta}_{-} \bar{\gamma}^{-1} \mathcal{S}_{+}}{\mathcal{S}_{-}} + \delta\Big[\frac{2}{3} \bar{\beta}_{-} - 2 \bar{\beta}_{-} \bar{\gamma}^{-1} + \frac{4 \mathcal{S}_{+}}{5 \mathcal{S}_{-}} -  \frac{2 \bar{\beta}_{+} \bar{\gamma}^{-1} \mathcal{S}_{+}}{\mathcal{S}_{-}}\Big]+ \frac{14}{15} \nu\Bigg\} \nn\\
&+ x^{3/2} \Bigg\{\frac{(2+\bar{\gamma})\pi}{2} - i\bigg(\frac{\left(2+\bar{\gamma}\right)\left(1+12 \ln(2)\right) }{12} +  \frac{1}{3} \zeta \left( \mathcal{S}_{+}^2+  \mathcal{S}_{-}^2 \right) 
 + \frac{4}{3}\zeta \mathcal{S}_-^2 \nu
 +  \frac{2}{3} \zeta \mathcal{S}_{-} \mathcal{S}_{+} \delta  \bigg)\Bigg\}   \,,
\end{align}
where the different post-Newtonian parameters are defined in our main paper \cite{Bernard:2022noq}.

\subsection{Flux and phase}

Using this formalism, we have also computed the flux for general and quasi-circular orbits to~1.5PN and the phase for dipolar-driven quasi-circular binaries to 1.5PN. These lengthy results are given in our main paper~\cite{Bernard:2022noq}. We are in agreement with the literature~\cite{Lang:2013fna,Lang:2014osa,Sennett:2016klh} (in which these quantities are compute to 1PN order), except for the 1PN flux in general orbits~\cite{Lang:2014osa} for which we find a discrepancy that could not be resolved, despite the fact that we agree on all the ST multipole moments separately. 


\section*{References}
\bibliography{ListeRef_ST}


%
%
%
%
%

\end{document}